\documentstyle[prl,aps,twoside,multicol]{revtex}

\newtheorem{theorem}{Theorem}
\newcommand{\bsig}{{\mbox{\boldmath $\sigma$}}}
\newcommand{\bI}{\bf \hat{I}}
\newcommand{\bH}{\bf \hat{H}}
\newcommand{\bF}{\hat{{\bf F}}}
\newcommand{\bG}{\hat{{\bf G}}}
\newcommand{\be}{\begin{equation}}
\newcommand{\ee}{\end{equation}}
\newcommand{\ba}{\begin{eqnarray}}
\newcommand{\ea}{\end{eqnarray}}
\newcommand{\ra}{\rangle}
\newcommand{\la}{\langle}

\begin{document}

\title{Decoherence Free Subspaces for Quantum Computation}

\author{
D. A. Lidar,$^{(a)}$
I. L. Chuang$^{(b)}$ and
K. B. Whaley$^{(a)}$
}

\address{
$(a)$ Department of Chemistry,
The University of California, Berkeley, CA 94720\\
$(b)$ IBM Almaden Research Center,
San Jose, CA 94120
}

\maketitle

\begin{abstract}
Decoherence in quantum computers is formulated within the Semigroup
approach. The error generators are identified with the generators of a Lie
algebra. This allows for a comprehensive description which includes as a
special case the frequently assumed spin-boson model. A generic condition is
presented for error-less quantum computation: decoherence-free subspaces are
spanned by those states which are annihilated by all the generators. It is
shown that these subspaces are stable to perturbations and moreover, that
universal quantum computation is possible within them.
\end{abstract}

\begin{multicols}{2}
\markboth{{Lidar, Chuang and Whaley}}{{To appear in {\it
Phys. Rev. Lett.}, 1998}}

\newpage

Decoherence remains the most important obstacle to the exploitation of the
speedup \cite{Shor:94Grover:97} promised by quantum computers. To this end
a remarkable theory of quantum error correction codes (QECC) has recently
been constructed \cite{shor:95calderbank:96Steane:96a}, in which a logical
quantum bit (qubit) is encoded in the larger Hilbert space of several
physical qubits.  This ``active'' error-{\em correction} approach builds on
the assumption that the most probable errors are those which occur
independently to a few qubits during a reasonable time interval.  However,
correlated errors, which affect many or all qubits, may also be likely in
some experimental realizations, particularly when qubits are physically
close (for example, nuclear spins in a molecule)\cite{Gershenfeld:97}.  
Such situations motivate the present study of an alternative
``passive'' error-{\em prevention} scheme, in which logical qubits are
encoded within subspaces which do not decohere because of reasons of
symmetry. The existence of such {\em decoherence-free} (DF) subspaces
has been shown by projection onto the symmetric subspace of multiple
copies of a quantum computer \cite{Barenco:97}, and by use of a
group-theoretic argument \cite{Zanardi:98}. Construction of
these subspaces has been performed explicitly for certain collective
error processes in the spin-boson model
\cite{Palma:96Duan:97PRLDuan:98Zanardi:97b,Zanardi:97c}. In this
Letter we formulate a general theory for decoherence in quantum
computation (QC) within the powerful semigroup approach
\cite{Lindblad:76,Alicki:87}, and show that this provides a
rigorous and comprehensive criterion for construction of DF subspaces
for an arbitrary Hamiltonian.\\
{\it The Semigroup Approach} ---. The dynamics of a quantum system
$S$
coupled to a bath
$B$
(which together form a closed system) evolves unitarily under the
Hamiltonian:
${\bH}_{SB}={\bH\otimes \bI}_{B}+{\bI}_{S}\otimes {\bH}_{B}+{\bH}_{I}$,
where
${\bH}$, ${\bH}_{B}$
and
${\bH}_{I}$
are the system, bath and interaction Hamiltonians, respectively.
$\bI$
is the identity operator. In the Semigroup approach one shows that under
the assumptions of (i) Markovian dynamics, (ii) ``complete
positivity'' \cite{Alicki:87}, and (iii) initial decoupling between
the system and bath \cite{Pechukas:94Pechukas+Alicki:95}, the
following master equation provides the most general form for the
evolution of the system density matrix
$\rho $:

\ba
\frac{\partial \rho }{\partial t} &=& {\tt L}\,[\rho ]
\equiv -\frac{i}{\hbar }[ {\bf H},\rho ]+{\tt L}_{D}\,[\rho ]
\label{eq:L} \\
{\tt L}_{D}\,[\rho ] &=&
\frac{1}{2}\sum_{\alpha ,\beta =1}^{M}
a_{\alpha \beta } {\tt L}_{{\bf F}_{\alpha },{\bf F}_{\beta }}[\rho]
\label{eq:L_D} \\
{\tt L}_{{\bf F}_{\alpha },{\bf F}_{\beta }}[\rho] &=&
[{\bf F}_{\alpha },\rho {\bf F}_{\beta }^{\dagger }] +
[{\bf F}_{\alpha }\rho ,{\bf F}_{\beta }^{\dagger }].
\ea
The commutator involving
${\bf H}$
is the ordinary, unitary, Heisenberg term. All the non-unitary,
decohering dynamics is accounted for by
${\tt L}_{D}$.
The time-independent {\em Hermitian} coefficient matrix
$A\equiv \{a_{\alpha \beta }\}$
contains the information about the physical decoherence parameters
(lifetimes, longitudinal or transverse relaxation times, and various
equilibrium parameters such as stationary polarization or
magnetization) \cite{Alicki:87}.\\
The $\{\bF_{\alpha }\}_{\alpha =0}^{M}$
($\bF_{0}=\bI $)
constitute a basis for the vector space of bounded operators acting on
${\cal H}$,
the $N$-dimensional system Hilbert space. This
operator-space may be restricted -- see the classification below.  As
such, the set 
$\{\bF_{\alpha }\}_{\alpha=1}^{M}$
forms an $M$-dimensional Lie algebra
${\cal L}$,
with an
$N\times N$ (generally $M \leq N^2-1$)
matrix {\em representation}
$\{{\bf F}_{\alpha }\}_{\alpha=1}^{M}$
appearing in Eq.~(\ref{eq:L_D}) (we omit the hat symbol for
matrices). Physically, the
$\{\bF_{\alpha }\}_{\alpha =1}^{M}$
describe the various decoherence processes: in the QC context they are
the {\em error generators}. They are often determined implicitly by the
interaction Hamiltonian:
\be
{\bH}_{I} =
\sum_{\alpha }{\bf \hat{F}}_{\alpha }\otimes {\bf \hat{B}}_{\alpha },
\label{eq:H_I}
\ee
where
$\{{\bf \hat{B}}_{\alpha }\}$
are bath operators (see Ref.~\cite{Kosloff:97} for examples).\\
{\it Decoherence of a Quantum Register ---.}
Consider a quantum computer made of
$K$
qubits. States in the corresponding
$N=2^{K}$-dimensional register Hilbert space
${\cal H}$ are tensor products of single qubit states
$|\varepsilon _{\kappa }\ra $,
$\varepsilon _{\kappa }=0,1$.
It is convenient to adopt the following classification of 
decoherence models of interest, in terms of the above Lie-algebraic scheme:
(i) ``{\em Total decoherence}:'' This provides the maximum possible
complexity of error generation, in which combined errors from any
number of qubits are generated. As is well known, {\em single}-qubit
errors can be fully described by the three Pauli matrices [i.e., the
defining representation of the Lie algebra
$ su(2)$].
Thus when
$|\varepsilon _{\kappa }\ra $
are the eigenstates of the
$\bsig_{\kappa }^{z}$
Pauli matrix, a single qubit can either undergo a phase-flip
($\bsig_{\kappa }^{z}$),
a bit-flip
($\bsig _{\kappa }^{x}$),
or both
($\bsig_{\kappa }^{y}$).
Taking into account also the possibility of no single-qubit error,
there are 4 possibilities per qubit, so that the maximal total number
of combined errors on
$K$
qubits is
$M=4^{K}-1$,
if we disregard the case of zero overall errors. The Lie algebra
$su(N)$
has
$N^{2}-1$
generators, so the corresponding
$M$
tensor products of Pauli matrices
$\{ \bF_{\alpha }\}$
form the defining representation of
${\cal L}=su(2^{K})$.
(ii) ``{\em Independent qubit decoherence}:'' In this, the ideal starting
point for QECC, we have the much simpler case of merely one independent
error per qubit, with all other qubits unaffected. There clearly are $3K$
such errors, each formed by taking the tensor product of a single Pauli
matrix on one qubit with the identity on all the rest. Since errors on
different qubits commute, this leads to a representation of the Lie algebra
${\cal L}=\oplus_{\kappa=1}^{K}su_{\kappa }(2)$.
(iii) ``{\em Collective decoherence}:'' One could also consider the
extreme case of all qubits undergoing the same decoherence process
simultaneously \cite{Zanardi:97c}, i.e., assuming full permutation
invariance of the qubits. There are then just 3 possible errors and
${\cal L}=su(2)$.
(iv) ``{\em Cluster decoherence}:'' Situations intermediate between
the above 3 cases follow when the register can be partitioned into
clusters
$k$
of
$K^{\prime }$
qubits, with collective decoherence taking place within each cluster,
but the clusters decohering independently. This leads to 
${\cal L}=\oplus _{k=1}^{K/K^{\prime }}su_{k}(2)$.
Lastly, a very interesting case (dealt with in detail below) arises
when a symmetry (e.g., permutation invariance) is broken {\em
perturbatively}.\\
{\it Conditions for Decoherence-free dynamics ---.}  Within the extremes
delineated by the above categorization, a particularly interesting question
is: what are necessary and sufficient conditions for the existence of a
generic DF subspace?  By generic (as opposed to general), we mean that one should (a) {\em avoid
fine-tuning of the noise parameters characterizing the decoherence
processes}, and (b) {\em avoid a dependence on initial conditions}. 
Suppose that
$\{|i\ra \}_{i=1}^{N_{0}}$
is a basis for an
$N_{0}$-dimensional
{\em invariant} DF subspace 
${\tilde{\cal H}}\subseteq {\cal H}$.
In this basis, we may express states as the density matrix
\be
	\tilde{\rho} = \sum_{i,j=1}^{N_{0}} \tilde{\rho}_{ij}|i\ra \la j| 
\,.
\label{eq:rho0}
\ee
Consider the action of the error generators on the basis states:
$\bF_{\alpha }|i\ra =\sum_{j=1}^{N_{0}}c_{ij}^{\alpha}|j\ra$.
The DF dynamics condition is
${\tt L}_{D}[\tilde{\rho}]=0$,
so that by Eq.~(\ref{eq:L}) the dynamics is purely unitary in the subspace
${\tilde{\cal H}}$.
Consider then Eq.~(\ref{eq:L_D}): condition (a) above implies that each
of the terms
${\tt L}_{{\bf F}_{\alpha },{\bf F}_{\beta }}[\tilde{\rho}]$
should vanish separately
$\forall \alpha ,\beta $.
A straightforward calculation yields:

\ba
{\tt L}_{{\bf F}_{\alpha },{\bf F}_{\beta }}[\tilde{\rho}] &=&
\sum_{ij,mn=1}^{N_{0}}
\tilde{\rho}_{ij}
\left(
2 c_{jm}^{\beta *} c_{in}^{\alpha }|n\ra \la m| \right. \nonumber \\
&-& \left. c_{mn}^{\beta *} c_{in}^{\alpha }|m\ra \la j|
-c_{jm}^{\beta *}c_{nm}^{\alpha }|i\ra \la n|
\right) .
\label{eq:x1}
\ea
To satisfy condition (b) above, each of the terms in parentheses must
vanish separately. This can only be achieved if there is just one
projection operator
$|n\ra \la m|$
in each term. The least restrictive choice leading to this is:
$c_{in}^{\alpha }=c^{\alpha}_i \delta_{in}$.
Eq.~(\ref{eq:x1}) then becomes:

\be
{\tt L}_{{\bf F}_{\alpha },{\bf F}_{\beta }}[\tilde{\rho}] =
\sum_{ij=1}^{N_{0}}
\tilde{\rho}_{ij}|i\ra \la j|
\left(
2 c^{\beta *}_j c^{\alpha }_i
-c^{\beta *}_i c^{\alpha }_i
-c^{\beta *}_j c^{\alpha }_j
\right) .
\label{eq:x2}
\ee
Assuming
$c^{\alpha}_i \neq 0$
then yields:
$\frac{c_{\alpha j}}{c_{\alpha i}}+\frac{c_{\beta i}^{\ast }}{c_{\beta
j}^{\ast }}=2$.
This has to hold in particular for
$\alpha =\beta $.
With
$z=c_{\alpha j}{/}c_{\alpha i}$,
we then obtain
$z+1/z^*=2$,
which has the unique solution
$z=1$.
This implies that
$c_{\alpha i}$
must be independent of
$i$
and therefore that
$\bF_{\alpha }|i\ra =c_{\alpha }|i\ra $,
$\forall \alpha $.
As a result we conclude that:
$[\bF_{\alpha },\bF_{\beta }]|i\ra =0$.
If $\cal L$ is {\em semisimple} (has no Abelian invariant subalgebra)
\cite{Cornwell:84II} then the commutator can be expressed in terms of
non-vanishing structure constants
$f_{\alpha,\beta}^{\gamma}$
of the Lie algebra:
$[\bF_{\alpha },\bF_{\beta }] =
\sum_{\gamma =1}^{M}f_{\alpha ,\beta }^{\gamma }\bF_{\gamma }$. We then arrive at the condition on the structure constants

\be
\sum_{\gamma=1}^M f_{\alpha,\beta}^{\gamma} c_{\gamma} = 0
\:\:\:\:\:
\forall \alpha,\beta .
\label{eq:f}
\ee
Now, it is known that the structure constants themselves define
the $M$-dimensional ``adjoint'' matrix representation of
${\cal L}$ 
\cite{Cornwell:84II}: 
$\left[ {\rm ad}(\bF_{\alpha }) \right]_{\gamma ,\beta } =
f_{\alpha ,\beta}^{\gamma }$.
Since the generators of the Lie algebra are linearly independent, so
must be the matrices of the adjoint representation. One can readily
show that this is inconsistent with Eq.~(\ref{eq:f}) unless all
$c_{\gamma }=0$.
We have thus proved \cite{Zanardi:98a}: 

\begin{theorem}
A necessary and sufficient condition for generic decoherence-free dynamics
$({\tt L}_{D}[\tilde{\rho}]=0)$
in a subspace
${\tilde{\cal H}}={\rm Span} [\{|i\ra \}_{i=1}^{N_{0}}]$
of the register Hilbert space, is that all basis states
$|i\ra $ are degenerate eigenstates of all the error generators
$\{\bF_{\alpha }\}$: $\bF_{\alpha }|i\ra =c_{\alpha }|i\ra $,
$\forall \alpha $; or, if ${\cal L}$ is semisimple, that all $|i\ra $
are annihilated by all $\{\bF_{\alpha }\}$: 
\be
{\bF}_{\alpha }\,|i\ra =0\quad \forall \alpha ,i .
\label{eq:DFD}
\ee
Equivalently, the DF subspace is spanned by those states transforming
according to the 1-dimensional irreducible representations (irreps) of
the Lie group with algebra
${\cal L}$.  Those states are {singlets}. 
The size of the DF code provided by this subspace is its dimension
$N_{0}$,
which can be used to further encode
$\log_2(N_0)$
logical qubits.
\end{theorem}

\noindent
Note also that by Eq.~(\ref{eq:H_I}):
${\bH}_{I}|i\ra \otimes |b\ra =0$,
where
$|b\ra $
is any bath state. Theorem 1 thus not only reduces the identification
of DF subspaces to a standard problem in representation theory of Lie
algebras, but also has the expected physical interpretation, namely
that the DF states are those that are annihilated by the interaction
Hamiltonian. (Note that this is only a {\em necessary} condition.)\\
{\it Effect of the system Hamiltonian} ---.
While
$\tilde{\rho}$,
by construction, is unaffected by the error
generators, the absence of decoherence may still be spoiled by the
system Hamiltonian itself. To see this explicitly, consider the {\it
mixed-state fidelity}:

\be
F(t)={\rm Tr}[\rho (0)\rho (t)]
= {\rm Tr}\left[\rho(0)\, \exp({\tt L\,}t)[\rho (0)]\right],
\ee
which is a natural measure of the decay of quantum coherence due to
coupling of the system with the environment. In ideal quantum computation,
one would like to have
$F(t)\!=\!1$,
corresponding to perfect, noiseless memory. In reality
$F(t)\!=\!1-\epsilon $,
$\epsilon \!>\!0$.
A formal power expansion yields: 

\be
F(t)~\!=\!\sum_{n=0}^{\infty }\frac{t^{n}}{n!}{\rm Tr}\left[ \rho (0)\left( 
{\tt L}\right) ^{n}[\rho (0)]\right] \equiv \!\sum_{n=0}^{\infty }\frac{1}{n!}
\left( \frac{t}{\tau _{n}}\right) ^{n}\text{,}  \label{eq:F-exp}
\ee
where the ``decoherence times'' are:
$\tau_{n} =
\left\{
{\rm Tr}\left[ \rho (0)\left( {\tt L}\right) ^{n}[\rho (0)] \right]
\right\} ^{-1/n}$.
In particular the first order decoherence rate is: 
\be
\frac{1}{\tau _{1}}={\rm Tr}\left[ \rho (0)~{\tt L}{\,}\rho (0)\right] \,.
\label{eq:tau_1}
\ee
Since
${\rm Tr}\left[ \rho [{\bH}, \, \rho ] \right] = 0$
(by cyclic permutation), it thus follows from Eq.~(\ref{eq:L}) that
$1/\tau _{1}=0$
for
$\tilde{\rho}$.
However, as is easily checked, generally
$1/\tau_{2}\neq 0$
because
${\bH}$
may cause transitions outside of
${\tilde{\cal H}}$.
Therefore, the {\it full dynamics} in
${\tilde{\cal H}}$,
including the effect of the system Hamiltonian, is DF to {\it first order}.\\
{\it Effect of Symmetry Breaking Perturbations} ---.
Suppose we have identified the DF subspace for the Lie algebra
${\cal L}$
underlying
${\tt L}_D$.
Let us consider the effect of adding new error generators
$\{ \bG_p\}_{p=1}^{P}$
which perturbatively break the symmetry, i.e., which do not belong to
${\cal L}$.
We assume that the
$\{\bG_{p}\}$
are due to an additional interaction Hamiltonian
${\bH}'_{I}$
which can be identified as appearing with a small parameter
$\epsilon $
in the full system-bath Hamiltonian:
${\bH}_{SB}={\bH}+{\bH}_{B}+{\bH}_{I}+\epsilon {\bH}'_{I}$.
Then the {\em new} terms added to
${\tt L}_D$
are:

\ba
{\tt L}'_{D}[\tilde{\rho}] &=&
\sum_{\alpha=1}^{M} \sum_{p=1}^{P}
\left(
a_{\alpha p} \,
{\tt L}_{{\bf F}_{\alpha },\epsilon {\bf G}_{p}}[\tilde{\rho}] +
a_{\alpha p}^* \,
{\tt L}_{\epsilon {\bf G}_{p},{\bf F}_{\alpha }}[\tilde{\rho}]
\right) \nonumber \\
&+& \sum_{p,q=1}^{P}
a_{pq} \,
{\tt L}_{\epsilon {\bf G}_{p},\epsilon{\bf G}_{q}}[\tilde{\rho}] .
\ea
Under the assumption
$\epsilon \ll 1$
we may neglect the last term since it is
$O(\epsilon^2)$.
As for the terms in the double sum,
${\bf F}_{\alpha } \tilde{\rho} =
\tilde{\rho}{\bf F}_{\alpha }^{\dagger }=0$
by Eqs.~(\ref{eq:rho0}) and (\ref{eq:DFD}). Expanding out the remaining
terms leaves: 

\be
{\tt L}_{D}^{\prime }[\tilde{\rho}]\approx \epsilon \sum_{\alpha
=1}^{M}\sum_{p=1}^{P}a_{\alpha ,p}\tilde{\rho}{\bf G}_{p}^{\dagger }{\bf F}
_{\alpha }+{\rm H.c.} 
\ee
While this will generally take the singlet states outside of
the DF subspace, this effect is also readily seen to be only of second
order, because the first-order decoherence time [Eq.~(\ref{eq:tau_1})]
is now given by:

\ba
\frac{1}{\tau _{1}} &=&
\epsilon \sum_{p=1}^{P}\left\{ a_{\alpha ,p}{\rm Tr}
\left[
\tilde{\rho}(0)\tilde{\rho}(0){\bf G}_{p}^{\dagger }{\bf F}_{\alpha }
\right] \right. \nonumber \\
&+& \left. a_{\alpha ,p}^* {\rm Tr}\left[
\tilde{\rho}(0){\bf F}_{\alpha
}^{\dagger }{\bf G}_{p}\tilde{\rho}(0)\right]
\right\} =0 \, ,
\ea
by cyclic permutation under the first trace. The higher
order decoherence-times,
$\tau_n$,
clearly involve
$\epsilon^n$
and can thus be made negligible. Therefore we have proved that {\bf
the DF subspace is {\em stable} to first order under a symmetry
breaking perturbation}.\\
This property is very promising from a quantum computational perspective,
since one should be able to apply standard QECC techniques to correct errors
which then occur within the DF subspace.  Of particular concern are errors
which take states out of the DF subspace; these are analogs of amplitude
damping errors, which abstractly model for example scattering and
spontaneous emission processes.  Such errors can be corrected by simple
codes\cite{Chuang:97aChuang:97bLeung:97}, for example, by taking the DF
singlet states as the computational basis states, and combining them into
QEC codewords. Provided that ${\bH}'_{I}$ causes independent errors on
different singlet 
states, we can conclude from the threshold
theorem\cite{Aharonov:96Knill:98,Preskill:97a} that as long as $\epsilon$ is
sufficiently small, the QECC encoding will render quantum computation within
$\tilde{{\cal H}}$ robust against these errors. Typical estimates of the
threshold error probability range from $10^{-6}$ to $10^{-3}$
\cite{Preskill:97a} and are extremely difficult to achieve in practice.  The
error probability is usually proportional to $\epsilon^2$.  However, within
$\tilde{{\cal H}}$, the error probability is reduced to $\epsilon^4$.  Thus,
QC within a DF subspace has potentially significant advantages.\\
{\it The dimension of DF subspaces: the size of codes} ---.
As shown in Ref. \cite{Zanardi:97c} for the spin-boson model, in
the limit of collective decoherence (i.e., when
${\cal L}=su(2)$)
the size of the DF subspace is:

\be
N_{0}\stackrel{K\gg 1}{\longrightarrow }K-\frac{3}{2}\log _{2}K .
\label{eq:CD}
\ee
The encoding efficiency
$N_{0}/K$,
is thus asymptotically unity. However, in the opposite limit of
independent qubit decoherence, 
${\cal L}=\oplus _{\kappa=1}^{K}su_{\kappa }(2)$,
which is addressed by QECC, there does {\em not} exist a DF subspace
\cite{DFD-comment2}. 
The size of the code obtained in intermediate cases of cluster decoherence
can be estimated from Eq.~(\ref{eq:CD}) by replacing
$K$
by
$K^{\prime }$
(the number of qubits per cluster), as long as
$K^{\prime }\lesssim K$.
However, the most interesting situation arises in the perturbative
scenario. Imagine a case of collective decoherence symmetry which is
perturbatively broken by small independent couplings between
individual qubits and the bath. As long as the symmetry-breaking
inhomogeneities are not too strong, we can conclude that, to first
order, the exponentially large DF subspace is still available.\\
{\it Universal Quantum Computation} ---.
Our discussion so far has centered on the preservation of quantum 
{\em memory}. To complete it we still need to show that universal
quantum computation can actually be performed in the DF
subspace. As is well known, the controlled-{\sc not} operation, together
with arbitrary single qubit rotations, can generate any unitary operation 
\cite{Barenco95a}. The corresponding unitary
operations are implemented by a driving Hamiltonian
${\bH}_{d}$,
which contains experimentally manipulable, time varying parameters,
together with the system Hamiltonian
${\bH}$. \\
We now give an example of universal 1- and 2-qubit operators acting on a
4-dimensional singlet subspace. Let
$|i\ra $,
$0\leq i\leq 3$
be singlet states. These 4 states span 2 encoded qubits
$|q_{1}\ra ,|q_{2}\ra $
where
$q_{1}q_{2}$
(with
$q_j=0,1$)
is the binary representation of
$i$.
A controlled-{\sc not} gate can be constructed from a Hamiltonian
represented in the encoded basis by the following combination of
projection operators:
${\bH}_{12}^{{\rm cnot}} =
c(t) \left[ |{\tt 11}\ra \la {\tt 10}| +
|{\tt 10}\ra \la {\tt 11}|\right]$.
Here
$c(t)$
is a time-dependent classical control parameter. Upon exponentiation
this yields the familiar conditional unitary operator form.
Single encoded-qubit rotations can be constructed from, e.g., 
${\bH}_{2}^{{\rm rot}} =
n_{0}(t)\left[ |{\tt 01}\ra \la {\tt 00}| +
|{\tt 00}\ra \la {\tt 01}| +
|{\tt 11}\ra \la {\tt 10}| +
|{\tt 10}\ra \la {\tt 11}|\right]$.
The generalization to larger singlet space systems is straightforward: one
constructs the appropriate projection operators on the {\em singlet}
states. By construction the resulting gates will leave the dynamics
DF. Thus, in principle, universal DF-QC is possible within the singlet
subspace. The main experimental challenge will involve implementation
of the corresponding operations on the {\em physical} qubits. In
addition, one should expect the actual implementation to involve some
of the amplitude damping errors discussed above, i.e., some ${\bH}_d$
operations will take the singlets out of the DF subspace. However, as
long as QECC is invoked, our previous arguments show that DF-QC is
still possible. \\
{\it Conclusions} ---.
It was shown how decoherence in QC can be described very generally in
terms of the Semigroup approach. The usual QC ``error generators''
were identified with the generators of a Lie algebra, whose identity
depends on the pertinent decoherence process. Without reference to a
specific system-bath interaction model, we derived {\em the generic
condition for DF subspaces}: these are spanned by those states which
are annihilated by all the error generators.  We showed further that
the DF subspaces are stable to first order under symmetry breaking
perturbations, which allowed us to extend their utility by application
of QECC. Finally, we showed that the DF subspaces support universal
quantum computation. 

{\it Acknowledgements} ---. This work was supported by NSF CHE-9616615
(to KBW) and by DARPA DAAG55-97-1-0341 (to ILC). We would like to
acknowledge helpful conversations with Drs. Paulo Zanardi, Robert
N. Cahn and Umesh Vazirani.

\end{multicols}


\begin{thebibliography}{10}
\bibitem{DFD-comment2}
Since the irreps of a direct sum algebra are given by the direct product of the
  irreps of the constituent algebras \cite{Cornwell:84II}, the representation
  realized for the $\{\bF_{\alpha }\}$ is therefore in this case already
  irreducible and (since it is a tensor product of $K$ $2\times 2$ matrices)
  $2^{K}$ dimensional. Thus it cannot contain any 1D irreps.
\end{thebibliography}
\end{document}